\documentclass[12pt,preprint]{aastex}
\newcommand{\Msun}{~M_\odot}

\newcommand{\kms}{\rm ~km~s^{-1}}

\newcommand{\ergs}{\rm ~erg~s^{-1}}

\newcommand{\ml}{~\Msun ~\rm yr^{-1}}

\begin{document}

\title{SHOCK BREAKOUT IN DENSE MASS LOSS: LUMINOUS SUPERNOVAE}
\author{Roger A. Chevalier and Christopher M. Irwin}
\affil{Department of Astronomy, University of Virginia, P.O. Box 400325, \\
Charlottesville, VA 22904-4325; rac5x@virginia.edu}

\begin{abstract}
We examine the case where a circumstellar medium around a supernova is sufficiently
opaque that a radiation dominated shock propagates in the circumstellar region.
The initial propagation of the shock front into the circumstellar region can be
approximated by a self-similar solution that determines the radiative energy in a shocked
shell; the eventual escape of this energy gives the maximum luminosity of the supernova.
If the circumstellar density  is described by
$\rho=Dr^{-2}$ out to a radius $R_w$, where $D$ is a constant, the properties of
the shock breakout radiation depend on $R_w$
and $R_d\equiv\kappa Dv_{sh}/c$, where $\kappa$ is the opacity
and $v_{sh}$ is the shock velocity.
If $R_w>R_d$, the rise to maximum light
begins at $\sim R_d/v_{sh}$; the duration of the 
rise is also $\sim R_d/v_{sh}$; the outer parts of the opaque medium are extended
and at low velocity
at the time of peak luminosity; and a dense shell forms whose continued interaction
with the dense mass loss gives a characteristic flatter portion of the declining light curve.
If $R_w<R_d$, the rise to maximum light
begins at $R_w/v_{sh}$; the duration of the 
rise is  $R_w^2/v_{sh}R_d$; the outer parts of the opaque medium are not extended
and are accelerated to high velocity by radiation pressure at the time of maximum 
luminosity; and a dense shell forms but does not affect the light curve near maximum.
We argue that SN 2006gy is an example of the first kind of event, while SN 2010gx and
related supernovae are examples of the second.

\end{abstract}


\section{INTRODUCTION}

Supernova shock breakouts from normal massive stars
typically give rise to X-ray/ultraviolet bursts with a timescale  $\la 10^3$ s
\citep{klein78,falk78,ensman92,matzner99,nakar10}.
The timescale is generally determined by $R_*/c$, where $R_*$ is the stellar
radius because the radiative diffusion time at breakout is less than the
light travel time across the star.
The largest red supergiants have a radius of $\sim 10^{14}$ cm \citep{levesque05}
so that the longest time of the breakout event is about 1 hour.
The short timescale of the bursts makes it difficult to detect them, other than
with a wide-field X-ray telescope.
The breakout detection of SN 2008D with {\it Swift} \citep{soderberg08}
was  fortunate.

The situation  changes if there is dense mass loss prior to the supernova
that creates an optically thick region.
Calculations of this case go back to early computer simulations of 
supernova light curves.
Model 5 of \cite{grassberg71} and Model B of \cite{falk77} include
a dense circumstellar shell with radius $\sim10^{15}$ cm that determines the properties of the
shock breakout.
The peak luminosity occurs on a timescale of $\sim15$ days in the model of \cite{grassberg71}.
\cite{grasberg87} considered a steady wind of limited duration just
before the supernova.
\cite{chugai04a} calculated a model for SN 1994W which allows for  dense mass loss around 
the supernova, leading to a peak luminosity at an age of $\sim20$ days.
A systematic numerical study of such events has recently been carried out by
\cite{moriya10}.
Because of the high surrounding density, the supernova can be especially bright and this type of model has been invoked for luminous supernovae.
In addition to SN 1994W \citep{chugai04a}, application 
of the dense mass loss model has been made to SN 2006gy 
\citep{smithmccray07}, PTF 09uj  \citep{ofek10}, and SN  2009kf  \citep{moriya10}, among others.

The aim here is to give an analytical description of the shock breakout process
for the case where a radiation dominated shock propagates into the
mass loss region. 
Progress on this front has been made by \cite{ofek10}.
The model is presented in \S~2 and compared to observations in \S~3.

\section{INTERACTION WITH DENSE MASS LOSS}
\label{diff}

We assume that the dense mass loss can be described by a steady wind acting for
some time $t_{ml}$ before the supernova explosion; the extent is $R_w=v_w t_{ml}$,
where $v_w$ is the wind velocity.
The actual case may be more complex, but is not well determined; the wind assumption has
been made in other work \citep{grasberg87,ofek10,moriya10}.
If the mass loss is in a steady wind,
the density  $\rho_w=\dot M/4\pi r^2v_w\equiv Dr^{-2}$ can be specified by a  density parameter, $D_*$,  scaled to a $\dot M=10^{-2}\ml$ and $v_w=10\kms$ wind
so that $\rho_w=5.0\times 10^{16}D_* r^{-2}$ in cgs units.
For typical supernova parameters, the explosion drives a radiation dominated
shock through the star.  
A radiation dominated shock in a uniform medium has a characteristic thickness described by
an optical depth $\tau_{sh}\approx c/v_{sh}$, where $v_{sh}$ is the shock velocity \citep{weaver76}. 
If the wind optical depth $\tau_w<c/v_{sh}$, then a radiation dominated shock
does not form in the wind and the radiation diffuses out through an
optically thick wind \citep[e.g.,][]{nakar10}.
In the case that $\tau_w>c/v_{sh}$, the radiation dominated shock front propagates
into the mass loss region.
This is the case considered here.

After the shock wave passes through the star and 
the supernova ejecta tend toward free expansion, a shocked layer develops
in the wind that is bounded by forward and reverse shock waves at $R_{fs}$ and 
$R_{rs}$, respectively.
The shock waves are initially energy conserving.
Once the forward shock front reaches a place where the diffusion time equals the
expansion time or, equivalently, $\tau_w\approx c/v_{sh}$, radiation diffusion
becomes important.
We have $\tau_w=(R_{fs}^{-1}-R_w^{-1})\kappa D$, so the diffusion condition can be
written as $R_d^{-1}\approx R_{fs}^{-1}-R_w^{-1}$, where 
$R_d\equiv \kappa Dv_{sh}/c=
5.7\times 10^{14}v_{sh4}kD_*$ cm, where $v_{sh4}$ is the shock velocity
in units of $10^4\kms$ and $k$ is the opacity $\kappa$
in units of 0.34 cm$^2$ g$^{-1}$.
Two cases are of interest: $R_w>R_d$ so that $R_{fs}\approx R_d$ when diffusion
becomes important, and $R_w<R_d$ so that $R_{fs}\approx R_w$ at that time.

In the wind, for diffusion over a distance $\Delta R$, the diffusion time is 
$t_d=\Delta R^2/\lambda c=\kappa\rho_w \Delta R^2/c$.
Taking $\Delta R\approx R$ and noting $\rho_w=Dr^{-2}$, we have
$t_d=\kappa D/c$, independent of radius \citep[see also][]{ofek10}.
If the extent of the optically thick region 
(to $R_{ph}=\min(R_w,\kappa D)$) is not much beyond $R_d$,
then $\kappa D/c$ provides an estimate of the diffusion time for the radiation
to escape.
The time is longer if $R_{ph}>R_d$.
If the diffusion is assumed to proceed in steps with size $\propto R$, the timescale
is increased by a term that is logarithmic in $R$.
For the approximate results presented here, we neglect the logarithmic dependence
and take
\begin{equation}
t_d=\kappa D/c=6.6 kD_*{\rm~day}.
\label{td}
\end{equation}

We now examine a specific model for expansion into the dense wind.
After the shock wave has expanded beyond the stellar radius,   
the ejecta tend toward free
expansion ($v=r/t$) in which there is a steep power law section on the outside and
a flat profile in the inner region \citep[e.g.,][]{matzner99}.
Here we make the approximation that $\rho_{in}\propto r^{-1}$ at velocities
$v<v_t$ and $\rho_{sn}\propto r^{-7}$ at $v>v_t$.
The outer profile is less steep than the profile inferred for the asymptotic
profile expected at high velocities; the wind interaction causes a 
flatter part of the density distribution to be relevant.
We then have $v_t=(2E/M_e)^{1/2}$ and $\rho_{sn}=Bt^{-3}(r/t)^{-7}$, where
$B=4E^2/3\pi M_e$ \citep{chevalier94}.
The interaction between this power law density distribution and a surrounding
wind, with density $\rho_w=Dr^{-2}$, can be described by a self-similar
solution with adiabatic index $\gamma=4/3$ \citep{chevalier83}, which is
appropriate for a radiation dominated gas.
The radius of the contact discontinuity is
\begin{equation}
R_{cd}=(0.23B/D)^{0.2}t^{0.8}=5.6\times 10^{14}E_{51}^{0.4}M_{e1}^{-0.2}D_*^{-0.2}t_1^{0.8} \rm{~cm},
\end{equation}
where $E_{51}$ is the explosion energy in units of $10^{51}$ ergs, $M_{e1}$ is the ejecta mass in units of $10\Msun$,  and $t_1$ is the age in units of 10 days.
The highest velocities occur just inside the reverse shock wave, where the velocity
is $v_{rs}=0.978R_{cd}/t=6.3\times 10^3E_{51}^{0.4}M_{e1}^{-0.2}D_*^{-0.2}t_1^{-0.2}\kms$.
The applicability of the solution requires that $v_t<v_{rs}$, or $E_{51}^{0.5}M_{e1}^{-1.5}D_* t_1<32$.

We initially assume that $R_d<R_w$.
At time $t_d$, 
\begin{equation}
R_{cd}=R_d=4.0\times 10^{14} k^{0.8}E_{51}^{0.4}M_{e1}^{-0.2}D_*^{0.6} {\rm~cm}.
\label{rd}
\end{equation}
The amount of energy in the interaction shell above some velocity $v_0$ is
\begin{equation}
E_{sh}=\int^{\infty}_{v_0t}{1\over 2}\rho_{sn}v^2 4\pi r^2 dr={4\over 3}{E^2\over M_ev_0^2}.
\end{equation}
The internal energy, which is in the form of radiation, is $E_{rad}=0.32E_{sh}$
\citep{chevalier83}.
Using $v_0=R_{cd}(t_d)/t_d$ at time $t_d$, we have
\begin{equation}
E_{rad}=0.44\times 10^{50}k^{0.4}E_{51}^{1.2}M_{e1}^{-0.6}D_*^{0.8} {\rm~ergs}
\end{equation}
for the radiative energy in the breakout shell.
The timescale for the radiation to escape is $t_d$, so the luminosity at shock breakout is
\begin{equation}
L_{b}\approx E_{rad}/t_d=7.6\times 10^{43}k^{-0.6}E_{51}^{1.2}M_{e1}^{-0.6}D_*^{-0.2}\ergs.
\end{equation}
Taking the radiation energy $E_{rad}$ and the shocked volume of $4\pi(R_{fs}^3-R_{rs}^3)/3$ at time $t_d$, where $R_{fs}$ and $R_{rs}$ are the forward and reverse shock radii, the mean radiation energy density gives a radiation temperature
\begin{equation}
T_{rad}=6.4\times 10^4 k^{-0.25}t_{d1}^{-0.25}=7.1\times 10^4  k^{-0.5}D_*^{-0.25}  {\rm~K},
\label{temp}
\end{equation}
where we have used $R_{fs}/R_{cd}=1.208$ and $R_{rs}/R_{cd}=0.978$ \citep{chevalier83},
and $t_{d1}$ is $t_d$ in units of 10 days.
Another estimate of $T_{rad}$ can be obtained from the forward shock condition on
the pressure: $aT^4/3=(6/7)\rho_w v_{fs}^2$ for a $\gamma=4/3$ gas, where $a$ is
the radiation constant \citep[see also][]{ofek10}.  
In this case, the temperature coefficient in equation (\ref{temp}) is 7\% higher;
the difference is due to the fact that there is a drop in the pressure behind the
forward shock front in the shocked region.
The results  assume a smooth wind with an $r^{-2}$ density profile, but
the temperature of the breakout shell is not strongly dependent on this assumption.
More generally, if $\Delta R$ is the size of the breakout region, the optical depth through
the region is $\rho \kappa \Delta R\approx c/v_{sh}$ and the shock crosses the shell 
in the diffusion time, or $\Delta R\approx v_{sh}t_d$.
The resulting density $\rho=c/\kappa v_{sh}^2 t_d$, when combined with the 
shock jump condition for the radiation pressure, gives a temperature comparable to
that in equation (\ref{temp}).

These results can be compared to the numerical results of \cite{moriya10}.
Taking their model s13w2r20m2e3 with $E_{51}=3$, $M_{e1}=1.3$, and $D_*=1$,
the breakout radius in our model is $R_d=5.9\times 10^{14}$ cm, which is inside
the outer radius of $2\times 10^{15}$ cm in the numerical model.
Substituting the parameters into equation (5), the radiated energy is 
$E_{rad}=1.4\times 10^{50}$ ergs, which is close to the $2.0\times 10^{50}$ ergs
found in the numerical model \citep{moriya10}.
The numerical result may be larger, in part, because  the shock wave
generates power in the more extended circumstellar medium.
We also examined the scaling of $E_{rad}$ with the parameters and found
reasonable agreement with the numerical results.
The scaling depends on the supernova density gradient, which is only
approximately treated here.

Estimates of the observable color temperature of the radiation require considerations
of whether radiation equilibrium is attained in the emitting region.
Following \cite{nakar10}, we define a thermal coupling coefficient
$\eta=n_{BB}/(t_d\dot n_{ph,ff}(T_{BB}))$, where $n_{BB}\approx aT^4_{BB}/3k_BT_{BB}$
is the photon number density in thermal equilibrium, $k_B$ is Boltzmann's constant, and $\dot n_{ph,ff}(T_{BB})=3.5\times 10^{36}\rho^2T^{-1/2}$
s$^{-1}$ cm$^{-3}$ is the production rate of photons by the free-free process.
If sufficient photons are produced to maintain the blackbody number density,
or $\eta\la 1$, thermal equilibrium is achieved.
Using equations (\ref{td}) and  (\ref{temp}) for $t_d$ and $T_{BB}$, and $\rho=7\rho_w(R_{fs})$
(taking into account the factor 7 compression in the shock wave), we estimate
$\eta$ for the breakout shell:
\begin{equation}
\eta_0\approx 0.4 k^{0.45}E_{51}^{1.6}M_{e1}^{-0.8}D_*^{-1.48}.
\end{equation}
For the standard parameters, the breakout shell is marginally in thermal equilibrium.
As the radiation propagates into the unshocked mass loss region, the lower density
results in a  deviation from thermal equilibrium.
The frequency dependence of the opacity can play a role \citep{moriya10} and
we do not treat the details of spectrum production here.

The loss of radiative energy from the shocked region results in the formation of a dense shell
at radius $R$,
as seen in numerical simulations \citep{grassberg71,falk77}.
The expansion of the shell into additional mass loss produces 
continuing power for the supernova,
$L=2\pi R^2 \rho_w (v_{fs}-v_w)^3$, where the wind velocity $v_w$ may be
affected by preshock radiative acceleration.
The simulations of \cite{moriya10} show some evidence for acceleration, but it 
is only significant near the breakout radius because of the $r^{-2}$ dependence
of the radiative flux, and we neglect it here.
The expansion of $R$ can be described by the thin shell approximation \citep{chevalier82},
yielding $R=0.94 R_{cd}$.
The resulting power is
\begin{equation}
L=7.1\times 10^{43}E_{51}^{1.2}M_{e1}^{-0.6}D_*^{0.4}t_1^{-0.6}\ergs.
\end{equation}
The magnitude of the luminosity is similar to that produced by the initial
breakout radiation, as is seen in numerical simulations \citep{grasberg87, moriya10}.
The  luminosity lasts until the shock wave at $R$ reaches the edge of the dense wind, $R_w$,
at $t_w=1.1E_{51}^{-0.5}M_{e1}^{0.25}D_*^{0.25}R_{w16}^{1.25}$ yr, where
$R_{w16}$ is in units of $10^{16}$ cm.
Fig. 1a illustrates the luminosity evolution with the late flattening from the
shell interaction.

In the case of shock breakout from a red supergiant, the shock front takes $\sim 1$ day
to traverse the star and the time for shock breakout is $\sim 10^3$ s \citep{klein78}.
The shock breakout timescale is much less than the time since explosion.
In the dense mass loss case considered here, the time for the shock front to move
to the breakout region is $\sim R_d/v_{sh}$, which is also the timescale for 
the breakout event.
This property of the luminosity evolution can be seen in simulations of
such events \citep{grassberg71,falk77,chugai04a,moriya10}.
The rise to maximum light can have complications due to variations
in the gas opacity.
At the high circumstellar densities considered here, it is likely that
the gas is initially neutral and that most of the opacity is due to dust in the presupernova environment.
The radiation dominated shock wave from the supernova has  a
precursor in the mass loss region that is expected to heat the circumstellar dust.
As the temperature rises through $1000-2000$ K, the dust evaporates, giving a decrease
in the opacity and the photospheric radius drops to where there is a sharp
gradient in the opacity as the gas becomes ionized.
In Type IIP supernovae, this property of the opacity causes a recombination wave to 
back into the expanding envelope with a constant photospheric temperature
$T\sim 5000-6000$ K.
Here, the process is inverted and the photosphere is expected to follow the ionization wave moving out through the dense circumstellar gas.
This phase of approximately constant 
temperature can be seen in simulations \citep{grassberg71}.
Once the circumstellar mass is ionized,
the photospheric expansion slows and the temperature rises.
The light curve rises fairly sharply due to the temperature rise until the
maximum temperature, and luminosity, is reached.

We now consider the case that $R_w<R_d$.
The shock breakout process begins at $t_b\approx R_w/v_{sh}$.
The rise time for the light curve is $t_r\approx\delta R/v_{sh}$,
where $\delta R$ is the distance in from $R_w$ across which the diffusion time 
equals the shock crossing time.
We thus have $(\delta R)^2\kappa DR_w^{-2}/c=\delta R/v_{sh}$,
leading to $t_{r}\approx (R_w/v_{sh})(R_w/R_d)$.
In this case the rise time can be considerably shorter that the time for
initial heating of the envelope (Fig. 1b).
Since $\delta R/R_w<1$, the region involved in shock breakout is not
radially extended, so the pressure of the escaping radiation can accelerate
the gas out to $R_w$ \citep{ensman94}. 
High gas velocities are expected around the time of maximum luminosity.
A dense shell forms when the radiation can escape, but does not produce
continuing high luminosity because the dense gas does not extend
far beyond the breakout point and it has been radiatively
accelerated.

For our specific supernova model,
the shock breakout begins when the forward shock ($R_{fs}$) reaches $R_w$,
which occurs at $t_w=16 R_{w15}^{1.25}E_{51}^{-0.5}M_{e1}^{0.25}D_*^{0.25}$ days,
where $R_{w15}$ is $R_w$ in  units of $10^{15}$ cm.
The free expansion velocity at the reverse shock is
$v_0=5.7\times 10^3 R_{w15}^{-0.25}E_{51}^{0.5}M_{e1}^{-0.25}D_*^{-0.25}\kms$,
which leads to a radiated energy of
$E_{rad}=0.65\times 10^{50}   R_{w15}^{0.5}\linebreak[2] E_{51}M_{e1}^{-0.5}D_*^{0.5}$ ergs.
The rise time for the light curve is
$t_r=t_wR_w/R_d=41 k^{-0.8} R_{w15}^{2.25}E_{51}^{-0.9}\linebreak[2] M_{e1}^{0.45}D_*^{-0.35}$ days,
so that $L\approx E_{rad}/t_r=1.8\times 10^{43}k^{0.8}R_{w15}^{-1.75}E_{51}^{1.9}M_{e1}^{-0.95}D_*^{0.85}\ergs$. 
The temperature in the shocked shell is like that given in the first part of
equation (\ref{temp}) except that $t_d$ is replaced by $t_w$.
The temperature is lowered in the escape out to the photosphere if 
there is sufficient photon production in this region \citep{nakar10}.

\section{COMPARISON WITH OBSERVATIONS}

The radiated energy from SN 2006gy from the initial optical rise to around the
peak is $\sim 1\times 10^{51}$ ergs
\citep{ofek07,smith07}, which makes it a good candidate for the
physical situation described here (with wind optical depth $\tau_w>c/v_{sh}$).
In this case, the observed rise time gives an estimate of $t_d$, the diffusion time.
The observations indicate a rise time of 60 days.
Using equation (\ref{td}), the indicated wind density is $D_*\approx 10$.
The observed peak luminosity has sensitivity to the supernova energy.
At maximum light the observed luminosity was $4\times 10^{44}\ergs$
\citep{smith10}, which implies $E_{51}\approx 3$ for the other standard parameters.
With these parameters, equation (\ref{rd}) gives $R_d=2.5\times
10^{15}M_{e1}^{-0.2}$ cm while the dense medium may extend to $\sim 1\times 10^{16}$ cm
\citep{smith10}, so that $R_w>R_d$.
We attribute the flattening of the observed light curve \citep{smith10} to the
continuing interaction in the extended region.
With $R_w= 10^{16}$ cm 
the implied circumstellar mass is $\sim 30\Msun$.
These parameters are close to those found by \cite{smithmccray07}
and \cite{smith10},
which is expected because the basic physical picture is similar in the two cases.
Depending on the ejecta mass, the condition that $v_t<v_{rs}$ may be violated,
but we do not expect the results to be strongly affected.

\cite{smith07,smith10} estimated an explosion date of 20 August, 2006 for SN 2006gy;
this time is just before the beginning of a rise in optical luminosity
from $0.01L_{max}$ to $L_{max}$, where $L_{max}$ is the maximum optical
luminosity.
In the shock breakout view, the sharp rise of optical radiation occurs after
a time $R_d/v_{sh}$ during which the shock is traveling in the optically thick
region (Fig. 1).
The explosion date is thus $\sim 60$ days earlier than the estimate of
\cite{smith07,smith10}, and
mean velocities of uniformly expanding ejecta are lower than the
estimates of Smith et al.

Another aspect of the rise to maximum in shock breakout is that the increasing
temperature as the shock breaks out is an important component of the
rising luminosity.
In the case of SN 2006gy, \cite{smith10} find a temperature $T\approx 11,000$ K on
days 65 and 71 (from Aug 20, 2006) (their Fig. 4), which is close to the time
of optical maximum light.
On day 36, \cite{smith10} estimate $T\approx 9,500$ K, using the same extinction
value as that used for other epochs (their Fig. 4).
However, they find that with an assumed larger extinction, a photosphere with
$T\approx 15,000$ K provides a better fit to the observed spectrum.
The higher temperature would bring the temperature evolution more in line with
that observed in other supernovae, i.e. a decreasing temperature with time,
and is advocated by \cite{smith10}.
However, in the shock breakout view, the increasing temperature is expected and
there is no need for a time dependent extinction.

In our model, the dense shell and shock wave are deep within the circumstellar
envelope, outside of which is the last equilibrium shell where the spectrum is formed.
The radius of this shell is not the blackbody radius $R_{bb}$ because of the
nonequilibrium conditions in the medium.  Outside of the shell is an
electron scattering region of moderate optical depth (between $c/v_{sh}$ and 1)
where the peaked H$\alpha$ profile with a broad base can be formed
\citep{chugai01,smith10}; the broadening is due to scattering in the thermal
gas as opposed to the Doppler effect.

An example of a different type of luminous supernova is SN 2010gx and related objects
\citep{pastorello10,quimby09}.
In this case, there is a sharp drop from peak luminosity without a flattening,
broad lines of O II are present at maximum light, and the narrow H or He lines
often observed in Type IIn supernovae are not present.
The evidence points to an explosion in a more compact circumstellar medium
for this class of objects.
The rise time, $\sim50$ days, and peak luminosity, $(3-4)\times 10^{44}\ergs$, are
comparable to the case of SN 2006gy, while the temperature is higher, $15,000$ K.
The rise time suggests that $R_w$ is not much less than $R_d$, so $R_w\approx R_d$.
The density is comparable to SN 2006gy.
The higher temperature in this case can be attributed to the lower opacity due to
the lack of H and He, and the smaller extent of the opaque circumstellar medium.

Although the case for dense mass loss years before the supernova appears good,
the cause of the mass loss is not known.
\cite{woosley07} suggested that SN 2006gy was the result of pair instability
eruption before the supernova.
Since the radiated energy produced in this case is $\sim 10^{50}$ ergs and
the radiated energy in observed luminous supernovae is $\ga 10^{51}$ ergs,
we have not specifically treated that case here, although if the energy were larger
similar physical arguments would presumably apply.
The dense mass loss is typically attributed to LBV (luminous blue variable) eruptions
\citep{smithmccray07}, although such eruptions are not understood in the context
of stellar evolution.
An eruption was observed 2 years before the explosion of SN 2006jc, which was a
H-poor, Type Ib supernova \citep{pastorello07,foley07}.
The mass loss in this case was too weak to produce the high density phenomena
discussed here, but it does show the possibility of a presupernova outburst,
even in the H-poor case.

{\it Note added:}  A model similar to that developed here was presented by
\cite{balberg11} and applied to the Type Ib SN 2008D.

\acknowledgments
We thank Claes Fransson for discussions.
This research was supported in part by NSF grant  AST-0807727.

\clearpage

\begin{figure}[!hbtp]   
\epsscale{1.0}
\plotone{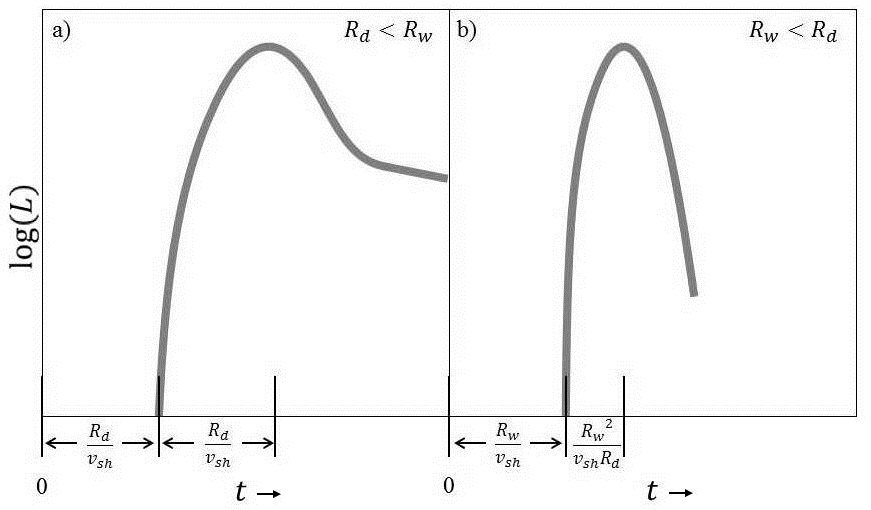}
\caption{Luminosity curves of two types of supernova interaction with dense mass
loss: a) wind extent $R_w$ greater than the characteristic diffusion radius $R_d$ and b) $R_w<R_d$.
In each case, there is a time from the explosion at $t=0$  to  the shock wave reaching a
place where the radiation can escape and the luminosity rises.
In the case $R_w>R_d$, there is a later, slower luminosity
decline due to continued interaction of the
shock wave (velocity $v_{sh}$) with slow wind material.
}
\end{figure}


\begin{thebibliography} {}

\bibitem[Balberg 
\& Loeb(2011)]{balberg11} Balberg, S., \& Loeb, A.\ 2011, MNRAS, submitted (arXiv:1101.1489)

\bibitem[Chevalier(1982)]{chevalier82} Chevalier, R.~A.\ 1982, 
\apj, 259, 302 

\bibitem[Chevalier(1983)]{chevalier83} Chevalier, R.~A.\ 1983, 
\apj, 272, 765 

\bibitem[Chevalier 
\& Fransson(1994)]{chevalier94} Chevalier, R.~A., \& Fransson, C.\ 1994, \apj, 420, 268 

\bibitem[Chugai(2001)]{chugai01} Chugai, N.~N.\ 2001, \mnras, 
326, 1448 

\bibitem[Chugai et al.(2004)]{chugai04a} Chugai, N.~N., et al.\ 
2004, \mnras, 352, 1213 

\bibitem[Ensman(1994)]{ensman94} Ensman, L.\ 1994, \apj, 424, 
275 

\bibitem[Ensman 
\& Burrows(1992)]{ensman92} Ensman, L., \& Burrows, A.\ 1992, \apj, 393, 742 

\bibitem[Falk(1978)]{falk78} Falk, S.~W.\ 1978, \apjl, 225, L133 

\bibitem[Falk 
\& Arnett(1977)]{falk77} Falk, S.~W., \& Arnett, W.~D.\ 1977, \apjs, 33, 515 

\bibitem[Foley et al.(2007)]{foley07} Foley, R.~J., Smith, N., 
Ganeshalingam, M., Li, W., Chornock, R., 
\& Filippenko, A.~V.\ 2007, \apjl, 657, L105    


\bibitem[Grassberg et 
al.(1971)]{grassberg71} Grassberg, E.~K., Imshennik, V.~S., \& Nadyozhin, D.~K.\ 1971, \apss, 10, 28 

\bibitem[Grasberg \& Nadyozhin(1987)]{grasberg87} Grasberg, E.~K., \& Nadyozhin, D.~K.\ 1987, \azh, 64, 1199 

\bibitem[Klein 
\& Chevalier(1978)]{klein78} Klein, R.~I., \& Chevalier, R.~A.\ 1978, \apjl, 223, L109 

\bibitem[Levesque et al.(2005)]{levesque05} Levesque, E.~M., 
Massey, P., Olsen, K.~A.~G., Plez, B., Josselin, E., Maeder, A., 
\& Meynet, G.\ 2005, \apj, 628, 973 

\bibitem[Matzner 
\& McKee(1999)]{matzner99} Matzner, C.~D., \& McKee, C.~F.\ 1999, \apj, 510, 379 

\bibitem[Moriya et al.(2010)]{moriya10} Moriya, T., Tominaga, 
N., Blinnikov, S.~I., Baklanov, P.~V., 
\& Sorokina, E.~I.\ 2010, \mnras, submitted (arXiv:1009.5799) 

\bibitem[Nakar 
\& Sari(2010)]{nakar10} Nakar, E., \& Sari, R.\ 2010, \apj, 725, 904 

\bibitem[Ofek et al.(2007)]{ofek07} Ofek, E.~O., et al.\ 2007, 
\apjl, 659, L13  

\bibitem[Ofek et al.(2010)]{ofek10} Ofek, E.~O., et al.\ 2010,  \apj, 724, 1396

\bibitem[Pastorello et al.(2007)]{pastorello07} Pastorello, A., et 
al.\ 2007, \nat, 447, 829    

\bibitem[Pastorello et al.(2010)]{pastorello10} Pastorello, A., et 
al.\ 2010, \apjl, 724, L16 

\bibitem[Quimby et al.(2009)]{quimby09} Quimby, R.~M., et al.\ 2009, \nat, submitted (arXiv:0910.0059) 


\bibitem[Smith 
\& McCray(2007)]{smithmccray07} Smith, N., \& McCray, R.\ 2007, \apjl, 671, L17   

\bibitem[Smith et al.(2007)]{smith07} Smith, N., et al.\ 2007, 
\apj, 666, 1116   


\bibitem[Smith et al.(2010)]{smith10} Smith, N., Chornock, R., 
Silverman, J.~M., Filippenko, A.~V., \& Foley, R.~J.\ 2010, \apj,    

\bibitem[Soderberg et al.(2008)]{soderberg08} Soderberg, A.~M., et 
al.\ 2008, \nat, 453, 469 

\bibitem[Weaver(1976)]{weaver76} Weaver, T.~A.\ 1976, \apjs, 32, 
233 


\bibitem[Woosley et al.(2007)]{woosley07} Woosley, S.~E., Blinnikov, S., \& Heger, A.\ 2007, \nat, 450, 390 


\end{thebibliography}
\end{document}